# An energy-based equilibrium contact angle boundary condition on jagged surfaces for phase-field methods

Florian Frank[a], Chen Liu[a], Alessio Scanziani[c], Faruk O. Alpak[b], Beatrice Riviere[a]

[a]*Rice University, Department of Computational and Applied Mathematics, 6100 Main Street, Houston, TX 77005, USA*
[b]*Shell Technology Center, 3333 Highway 6 South, Houston, TX 77082, USA*
[c]*Imperial College London, Department of Earth Science and Engineering, London SW7 2AZ, UK*

**Abstract**

We consider an energy-based boundary condition to impose an equilibrium wetting angle for the Cahn–Hilliard–Navier–Stokes phase-field model on voxel-set-type computational domains. These domains typically stem from the μCT imaging of porous rock and approximate a (on μm scale) smooth domain with a certain resolution. Planar surfaces that are perpendicular to the main axes are naturally approximated by a layer of voxels. However, planar surfaces in any other directions and curved surfaces yield a jagged/rough surface approximation by voxels. For the standard Cahn–Hilliard formulation, where the contact angle between the diffuse interface and the domain boundary (fluid–solid interface / wall) is 90°, jagged surfaces have no impact on the contact angle. However, a prescribed contact angle smaller or larger than 90° on jagged voxel surfaces is amplified in either direction. As a remedy, we propose the introduction of surface energy correction factors for each fluid–solid voxel face that counterbalance the difference of the voxel-set surface area with the underlying smooth one. The discretization of the model equations is performed with the discontinuous Galerkin method, however, the presented semi-analytical approach of correcting the surface energy is equally applicable to other direct numerical methods such as finite elements, finite volumes, or finite differences, since the correction factors appear in the strong formulation of the model.

## 1. Introduction

Digital Rock software is a critical tool for the oil and gas industry for developing a rigorous understanding of subsurface flow and transport and for the computation of rock properties such as absolute and relative permeabilities. Models of rock are obtained by modern micro-tomography (μCT) describing the density distribution of small rock samples on μm to mm scale by means of X-rays, 3D image processing and segmentation [1]. While Digital Rock simulations for poroelasticity sometimes compute a smooth triangulation of the fluid–solid interface based on the given voxel density data, fluid flow simulations usually operate directly on the image data, which in this case consist of a derived set of binary voxels, each of which is either associated with the pore space of the rock sample or with its solid matrix [2, 3]. A popular method to model two-phase flow for binary mixtures is the phase-field / diffuse interface approach, which is based on free energy minimization of the mixture [4, 5, 6, 7]. An often used phase-field model consists of the Cahn–Hilliard (CH) equation coupled to the Navier–Stokes (NS) equation by a velocity-dependent advection term in the former equation and a surface tension force in the latter one [8, 9, 10]. Solutions of the Cahn–Hilliard–Navier–Stokes (CHNS) system consist of the phase distribution, the mixture velocity and pressure, where the diffuse two-phase liquid–liquid interface is a smooth transition region of small, finite thickness.

Modeling wettability and contact lines between the fluid–fluid interface and the solid surface is an important task within Digital Rock workflows, as they influence the displacement of the fluids through the pore space. Wettability properties are reflected by the contact angle between the diffuse interface and the solid wall. There are two known general approaches for imposing a contact angle in phase-field methods: The geometric approach proposed by Ding and Spelt [11] and extended to non-equilibrium systems by Alpak et al. [12] is based on interface normal and tangent computations as functions of the mixture gradient, which is approximated by finite difference stencils (see [13] for a comparison study). Jacqmin [14, 15] postulated a surface-energy functional that can be expressed as a function of the contact angle and the mixture composition at the fluid–solid interface. Minimization of this functional leads to a boundary condition that controls the equilibrium contact angle—the angle between the center of the diffuse interface and solid wall at thermodynamical equilibrium. The closed-form expression of this energy was later expressed in terms of the equilibrium contact angle and the interfacial tension [16, 17]. Even though Ding and Spelt show the equivalence of the two approaches, in our numerical experiments the energy-based approach turned out to be more robust, in particular, since it supports the extreme contact angles of 0° and 180°.

The energy-based contact angle boundary condition is a valid model for smooth surfaces, which is attested by the successful numerical implementation in many articles [16, 17, 18]. In the context of Digital Rock, however, where jagged surfaces naturally arise from the underlying image data stemming from the µCT imaging of porous rock, this boundary condition is not suitable in its original formulation. Numerical simulations show that contact angles away from 90° are amplified on jagged surfaces, cf. Sec. 5. As a remedy, we propose a modified contact angle boundary condition that includes surface energy correction factors for each fluid–solid voxel face. These correction factors counterbalance the difference of the voxel-set surface area with one underlying smooth one. This is the main contribution of this paper. There are few papers in the literature on this topic. We mention the work of [19, 20] which smoothens the solid wall normal vectors in the context of a volume-of-fluid-based finite volume method to handle jagged surfaces. To the best of our knowledge, our approach is the first one proposed for phase-field/diffuse interface methods.

In the next section, the CHNS model including contact angle boundary condition is introduced. Our proposed method for imposing contact angle on jagged surfaces is described in Sec. 3 and Sec. 4. Section 5 deals with a series of benchmark scenarios with increasing geometric complexity to validate the proposed contact angle boundary condition on jagged surfaces.

## 2. Model problem

Let $J := (0, t_{\text{end}})$ denote a time interval with end time $t_{\text{end}} > 0$. We consider an incompressible mixture of two immiscible fluids inside the pore space $\Omega \subset \mathbb{R}^3$ of a porous medium at constant temperature in the context of the phase-field approach.

### 2.1. Mixture composition and fluid free energy

The mixture composition is described by an order parameter $C = C(t, \boldsymbol{x})$, which we choose as the difference of volume or mass fractions $C_A, C_B$ of either fluid, i.e. $C := C_A - C_B = 2 C_A - 1$, where $C(t, \boldsymbol{x}) \in [-1, 1]$. Cahn and Hilliard [6, 7] proposed that the free energy $F_f$ of the fluid can be approximated by

$$F_f(C, \boldsymbol{\nabla} C) := \int_\Omega \left( \beta \, \Psi(C) + \frac{\alpha}{2} \, \boldsymbol{\nabla} C \cdot \boldsymbol{\nabla} C \right), \quad (1)$$

where $\beta \, \Psi$ is called *homogeneous free energy density* corresponding to the free energy of a unit volume of homogeneous composition $C$. A frequently used expression for $\Psi$ is the Ginzburg–Landau potential [21] $\Psi(C) := \frac{1}{4}(C^2 - 1)^2$ The function $\Phi$ has the shape of a double-well potential with two minima at the bulk compositions, i.e., $C \in \{-1, 1\}$, provided that the constant system temperature is below the critical temperature of the mixture (i.e. where a homogeneous mixture becomes energetically unfavorable). While the homogeneous free energy term is responsible to decomposition of the mixture, the second term in the integrand of (1)—sometimes called *interfacial* or *gradient energy density*—is responsible for the diffuse nature of the interface between fluid phases. The factors $\alpha$ and $\beta$ are expressed in terms of the surface tension between the fluid phases $\sigma_{AB}$ and the interface parameter $\epsilon$ ($\epsilon$ is often called "interface width", however, the interface width is approximately $4 \epsilon$, cf. [22]):

$$\alpha := \frac{3}{2\sqrt{2}} \sigma_{AB} \epsilon, \quad \beta := \frac{3}{2\sqrt{2}} \frac{\sigma_{AB}}{\epsilon}. \quad (2)$$

The variational/functional derivative of the fluid free energy with respect to $C$,

$$\phi := \delta F_f = \beta \, \Psi'(C) - \alpha \, \Delta C \quad \text{in } J \times \Omega, \quad (3)$$

is usually called chemical potential, since the gradient of which induces an energy exchange (in form of a flux), even though its unit is not joule per mole but joule per cubic meter. The corresponding flux reads $\boldsymbol{j}_1 := -M \, \boldsymbol{\nabla} \phi$, where the phenomenological coefficient $M > 0$ is called *mobility*, which is assumed constant here. Assuming that the mixture is subjected to a solenoidal velocity $\boldsymbol{u} : J \times \Omega \to \mathbb{R}^3$, the total flux also includes advection: $\boldsymbol{j}_2 := \boldsymbol{u} \, C$. With the conservation equation for $C$, it follows

$$\partial_t C + \boldsymbol{\nabla} \cdot (\boldsymbol{j}_1 + \boldsymbol{j}_2) = \partial_t C - M \, \Delta \phi + \boldsymbol{u} \cdot \boldsymbol{\nabla} C = 0 \text{ in } J \times \Omega. \quad (4)$$

The system (3, 4) is known as the advective CH equation characterizing phase segregation, which is the alignment of the mixture into spatial domains predominated by one of the two components. Using the definition of $\Phi$ above, the interface profile between the phases has the shape of a hyperbolic tangent function. Subjected to the boundary conditions $\nabla C \cdot \boldsymbol{n} = \nabla \phi \cdot \boldsymbol{n} = 0$, where $\boldsymbol{n}$ denotes the unit normal on $\partial \Omega$ exterior to $\Omega$, the system dissipates free energy $F_\text{f}$ in time if the velocity is set to zero.

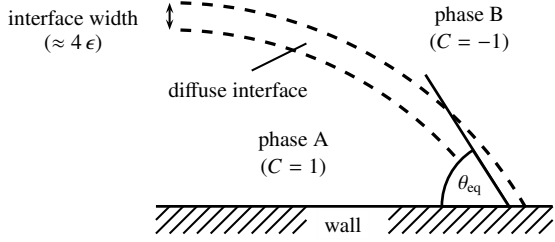

Figure 1: Illustration of the contact angle $\theta_\text{eq}$. The wetting property is called *neutral* for $\theta_\text{eq} = 90°$, *hydrophobic* for $\theta_\text{eq} > 90°$, *superhydrophobic* for $\theta_\text{eq} = 180°$ (droplet repels from surface), *hydrophilic* for $\theta_\text{eq} < 90°$, and *superhydrophilic* for $\theta_\text{eq} = 0°$ (wall is completely wetted).

## 2.2. Wall energy and contact angle

The angle between the (center of) the diffuse interface and the wall toward phase A ($C = 1$) is called *contact* or *wetting angle*, and is denoted by $\theta_\text{eq}$, with $\theta_\text{eq} \in [0°, 180°]$, cf. Fig. 1. At equilibrium state, the angle $\theta_\text{eq}$ depends on the surface tension $\sigma_\text{As}$ between phase A and the solid phase, on the surface tension $\sigma_\text{Bs}$ between phase B and the solid phase, and on the the surface tension $\sigma_\text{AB}$ between phase A and B [23]. The relationship between $\theta_\text{eq}$ and $\sigma_\text{AB}$ is given by Young's equation:

$$\cos(\theta_\text{eq}) = \frac{\sigma_\text{Bs} - \sigma_\text{As}}{\sigma_\text{AB}}. \quad (5)$$

Based on Jacqmin's postulated surface-energy functional [14, 15], the mixture composition at the fluid–solid interface is associated with an energy

$$F_\text{w}(C) := \int_{\partial\Omega} ((\sigma_\text{Bs} - \sigma_\text{As}) g(C) + \sigma_\text{As}), \quad (6)$$

where $g(C) := \frac{1}{4}(C^3 - 3C + 2)$ is a function that blends the surface tension $\sigma_\text{As}$ smoothly into $\sigma_\text{Bs}$ across the diffuse interface [17, 16]. The closed-form expression of $g$ depends on that of $\Psi$, cf. [24]. For the case of $\boldsymbol{u} = \boldsymbol{0}$, the CH equation (3), (4) and the equilibrium contact angle boundary condition

$$\alpha \nabla C \cdot \boldsymbol{n} = -\sigma_\text{AB} \cos(\theta_\text{eq}) g'(C) \quad \text{on } J \times \partial\Omega \quad (7)$$

can be derived by the variation of the functional $F_\text{f} + F_\text{w}$, cf. [24, 25]. Equation (7) has no influence on mass balance and it reduces to the standard condition $\nabla C \cdot \boldsymbol{n} = 0$ for $\theta_\text{eq} = 90°$. Note that (7) forces an angle equal to $\theta_\text{eq}$ at every time point $t$ in $J$.

## 2.3. Fluid flow

The mixture is assumed incompressible, hence, the velocity $\boldsymbol{u}$ is subjected to the mass balance equation

$$\nabla \cdot \boldsymbol{u} = 0 \quad \text{in } J \times \Omega. \quad (8)$$

Conservation of momentum yields the NS equation

$$\rho (\partial_t \boldsymbol{u} + \boldsymbol{u} \cdot \nabla \boldsymbol{u}) = -\nabla p + \mu \Delta \boldsymbol{u} - C \nabla \phi \quad \text{in } J \times \Omega, \quad (9)$$

which includes the term $C \nabla \phi$ being a force density due to surface tension across the diffuse interface. For the ease of presentation, the mixture density $\rho$, and the mixture viscosity $\mu$ are assumed to be constant (in particular, independent of the mixture composition).

## 2.4. CHNS system and energy law

Given the initial data $\boldsymbol{u}^0 : \Omega \to \mathbb{R}^3$, $C^0 : \Omega \to [-1, 1]$, the *CHNS problem* consists of seeking unknowns $C : \overline{J} \times \overline{\Omega} \to \mathbb{R}$, $\boldsymbol{u} : \overline{J} \times \overline{\Omega} \to \mathbb{R}^3$ such that the equations (3), (4), (8), (9) subjected to initial conditions $C = C^0$, $\boldsymbol{u} = \boldsymbol{u}^0$ on $\{0\} \times \Omega$ and to boundary conditions (7) and $\boldsymbol{u} = \boldsymbol{0}$, $\nabla \phi \cdot \boldsymbol{n} = 0$ on $J \times \partial\Omega$ are satisfied. The latter two boundary conditions suppress the flux $(\boldsymbol{j}_1 + \boldsymbol{j}_2)$ across $\partial\Omega$.

The total energy of the mixture is given by

$$F := F_\text{f} + F_\text{w} + F_\text{k} \quad (10)$$

with $F_\text{k}(\boldsymbol{u}) := \frac{\rho}{2} \int_\Omega \boldsymbol{u} \cdot \boldsymbol{u}$ being the kinetic energy. Solutions $(C, \boldsymbol{u})$ of the CHNS system satisfy the following *energy dissipation law*:

**Proposition 1.** *The total energy is non-increasing in time, i. e., $\forall t \in J$, $\text{d}_t F(t) \leq 0$. In particular,*

$$\text{d}_t F = -\int_\Omega (\mu \nabla \boldsymbol{u} : \nabla \boldsymbol{u} + M \nabla \phi \cdot \nabla \phi).$$

This law states that the decrease in time of the total free energy of the system equals the dissipation due to viscosity and the dissipation due to diffusion in the bulk. The proof of Prop. 1 is found in Sec. 7. For nondimensionalization of the system, see, e. g., [26, 27]. We skip this step in this article since the introduced energy correction for jagged surfaces is independent of the scaling.

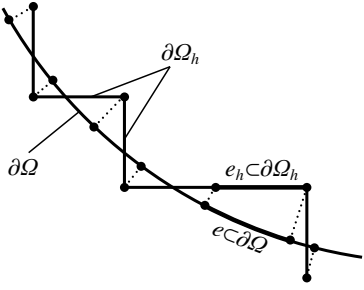

Figure 2: Sketch of the smooth boundary $\partial\Omega$ and its corresponding voxel surface $\partial\Omega_h$. Every face $e_h$ of $\partial\Omega_h$ defines a patch $e$ on $\partial\Omega$ by orthogonal projection.

## 3. Boundary condition on jagged surfaces

Recall that $\Omega_h$ denotes the union of voxels that approximates some smooth domain $\Omega$ (which is not known in the context of porous imaging). The contact angle boundary condition (7) describes a surface energy density (with unit joules per meter square) related to a *smooth* surface $\partial\Omega$. If this boundary condition is applied to the jagged surface $\partial\Omega_h$ instead, the contact angle is spuriously amplified, which is illustrated in Sec. 5.

Let us momentary assume that we know the smooth surface $\partial\Omega$ of the domain $\Omega$ of which $\Omega_h$ is the imaging data. Heuristically motivated, we propose that the prescribed energy of $\partial\Omega$ must be equal to the energy of the voxel surface $\partial\Omega_h$, cf. Fig. 2. An even stronger claim is asking for this equality *locally*, i.e., for each face $e_h \subset \partial\Omega_h$, the prescribed energy on $e_h$ must be equal to a corresponding patch $e \subset \partial\Omega$. We propose to define these patches on $\partial\Omega$ by orthogonal projection from $e_h \subset \partial\Omega_h$ onto $\partial\Omega$. In doing so, the patches $e$ on $\partial\Omega$ form a non-overlapping partition of $\partial\Omega$.

With the considerations above, the boundary condition (7) for the jagged voxel surface has the form

$$\alpha \, \boldsymbol{\nabla} C_h \cdot \boldsymbol{n} = -\delta \, \sigma_{\text{AB}} \, \cos(\theta_{\text{eq}}) \, g'(C_h) \quad \text{on } \partial\Omega_h \, , \quad (11)$$

where $\delta$ is a face-wise positive constant and where $C_h$ is a discrete approximation of $C$. In order to find a closed-form expression for $\delta$, we assume that the order parameter and its discrete approximation are constant and equal to each other in the vicinity of one face $e_h$, more precisely, $C = C_h = $ const. By assumption, the energy on $e_h \subset \partial\Omega_h$ and on its corresponding patch $e \subset \partial\Omega$ is the same:

$$\int_e \sigma_{\text{AB}} \, \cos(\theta_{\text{eq}}) \, g'(C) = \int_{e_h} \delta \, \sigma_{\text{AB}} \, \cos(\theta_{\text{eq}}) \, g'(C_h)$$
$$\Leftrightarrow \quad \int_e 1 = \int_{e_h} \delta$$

and thus we obtain the formula

$$\delta|_{e_h} := \frac{|e|}{|e_h|} \quad \forall \, e_h \in \partial\Omega_h \, . \quad (12)$$

In order to use the boundary condition (11), for each face $e_h \in \partial\Omega_h$, the corresponding patch $e \in \partial\Omega$ must be known in order to compute $\delta$ using (12). This issue will be addressed in Sec. 4.

Boundary condition (11) is also valid for non-advective CH systems (without velocity), since stationary solutions of the CH equation are also stationary solutions of the CHNS system. In fact, the surface sension force-density term in (9) vanishes at equilibrium state. It holds $d_t F = 0$ in equilibrium, and with Proposition 1 it follows that $\phi$ and $\boldsymbol{u}$ are constant in $\Omega$. With the prescribed no-slip condition for $\boldsymbol{u}$, it holds $\boldsymbol{u} = \boldsymbol{0}$ in $\Omega$ at equilibrium state.

## 4. Computation of the energy correction factors

In Sec. 3, we assumed that the smooth surface $\partial\Omega$ is given. However, in practical applications, the only available information about the domain is given by the image data $\Omega_h$. We propose a simple heuristic approach to reconstruct a smooth surface *locally*: For each face $e_h \in \partial\Omega_h$, we mark the corresponding solid voxel and surrounding solid voxels that contribute to the solid–fluid interface $\partial\Omega_h$. We subsequently fit a polynomial surface through the centers of the marked voxels, cf. Fig. 3. The corresponding patches $e$ to $e_h$ are then determined by finding the four points on the polynomial surface that have the least distance to the vertices of $e_h$, cf. Fig. 2. The area $|e|$ can than be easily computed by parameterizing the polynomial surface with a grid. We note that this local approach is computationally cheap. However it has the disadvantage that the set of computed patches $\{e\}$ may not be a partition of $\partial\Omega$. A more appropriate and more costly approach is to use a *global* surface reconstruction approach that takes the whole voxel data set of $\Omega_h$ into account, cf. [28, 29]. In the context of pore-scale modeling, such an approach ideally preserves the connectivity of the porous domain and the volume of the pore space.

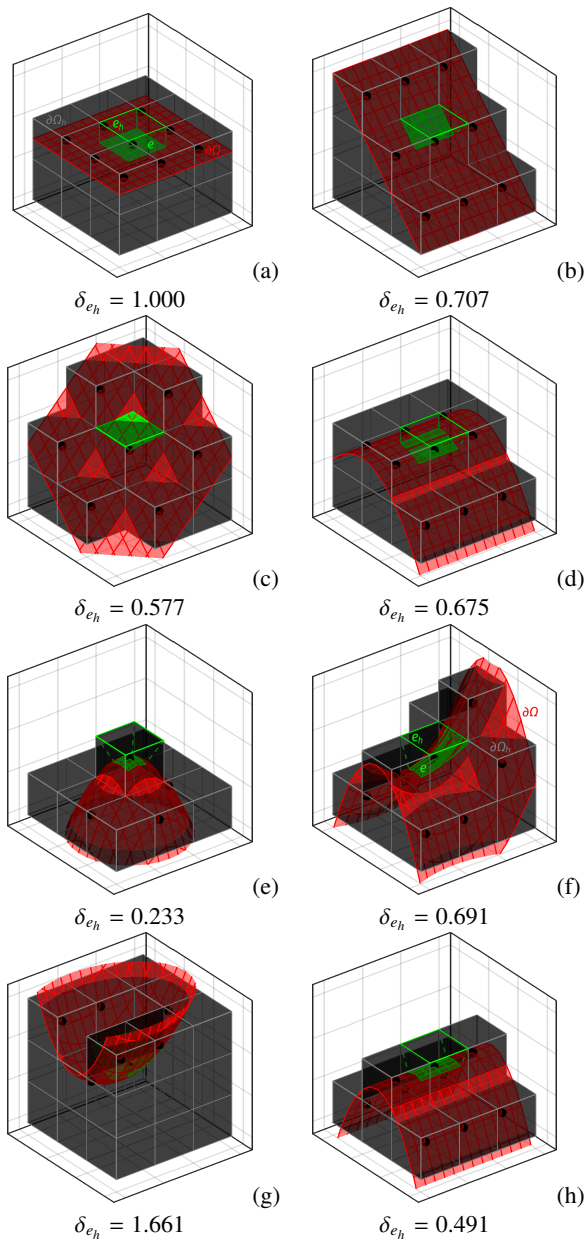

Figure 3: Local smooth surface approximation $\partial\Omega$ *(red)* in the vicinity of a face $e_h$ *(light green)* at the fluid–solid interface *(solid voxels in gray)*. The patch $e \in \partial\Omega$ is identified by an orthogonal projection of $e_h$ onto $\partial\Omega$. The computed energy correction factor $\delta_{e_h}$, cf. (12), corresponding to $e_h$ is indicated below each example geometry.

## 5. Numerical results

We use a semi-implicit convex–concave splitting [30, 31] for the temporal discretization of the CH subsystem (3), (4), while the NS subsystem (9), (8) is linearized by a Picard splitting and treated by a first order incremental pressure-correction scheme in rotational form [32]. The nonlinear boundary condition (11) is treated explicitly in time. The spatial discretization of the differential operators is obtained by applying the interior penalty discontinuous Galerkin method [33] with a hierarchical basis on each voxel. An efficient solution strategy for this type of discretization is proposed in [34]. For (3), (4), the scheme reduces to finite volumes for zeroth order polynomials. Details are given in [35, 36], where the the scheme and its implementation were verified by showing optimal convergence of the approximation error. Our numerical tests indicate that the time-space discrete scheme dissipates the total discrete energy in time.

The parameter $0 < \epsilon \ll 1$ is typically chosen as small as possible to generate a small interface width. Equilibrium profiles of $C$ are obtained by solving $\phi(C) = 0$ for $C$, which yields an interface width of approximately $4\epsilon$ in one space dimension, cf. [22] and references therein for suitable definitions the interface width. Simulation studies have shown that this estimate also holds in three dimensions, and that in numerical simulations $\epsilon$ has to be limited from below by the mesh size $h$ ($h$ equals the edge length of one voxel). Having at least four elements across the interface is sufficient in our numerical scheme, which requires that $\epsilon \geq h$. Note that the contact angle $\theta_{eq}$ is independent of $\epsilon$ [26].

This section introduces three numerical scenarios with increasing geometrical complexity of $\Omega_h$, in which the equilibrium contact angles $\theta_{eq}$ of droplets sitting on the surface $\partial\Omega_h$ are investigated. In Sec. 5.1, the droplet sits on a plane approximated by voxels. For this special case, the radius of the intersection disc between droplet and solid can be computed analytically, and we show that the proposed boundary condition (11) yields the correct contact angle on jagged surfaces. This section, in which we use an elementwise linear approximation of the CHNS system, also demonstrates that its equilibrium mixture composition coincides with the one of the CH subsystem (3), (4). In Sec. 5.2, a droplet is put onto a spherical surface with a more complex local geometry at the three-phase contact line. Section 5.3 deals with a droplet in a pore cavity of a Castlegate Sandstone sample, obtained by µCT scanning. In Sec. 5.2 and Sec. 5.3, we use an elementwise constant finite volume approximation of (3), (4).

### 5.1. Droplet on a flat plate with analytical solution
#### 5.1.1. General setup

Motivated by the work of [19], we consider a computational domain $\Omega_h \subset [0, 1]^3$ consisting of a set of voxels overlying a half ball. This domain may be rotated around the point $(0.5, 0.5, 0.5)$ in order to create

braic expression

$$r_{sh} = \sin(\theta_{eq}) \left( \frac{3\,V}{(\pi\,(1 - \cos(\theta_{eq}))^2(2 + \cos(\theta_{eq})))} \right)^{1/3}. \tag{13}$$

Equation (13) is a good approximation for the radius in the diffuse interface case provided that the interface is sufficiently small.

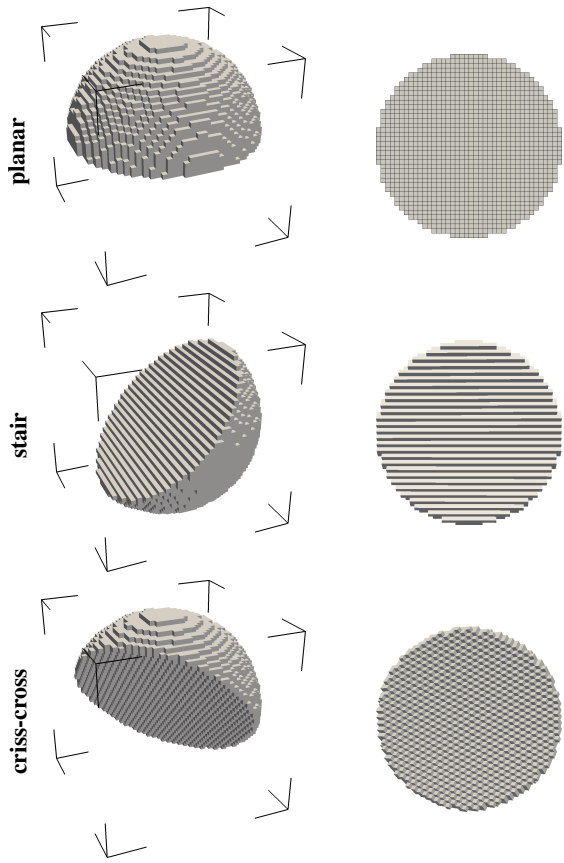

Figure 4: Computational domain $\Omega_h \subset [0, 1]^3$ (pore space, *gray*) consisting of voxels of size $h=1/40$ overlaying a half ball without rotation *(top)*, rotated by 45° in *x* direction *(center)*, and rotated by 45° in both *x* and *y* direction *(bottom)*.

different types of jagged voxel approximations of the circular bottom, cf. Fig. 4. The initial order parameter $C^0$ is chosen equal to 1 in a rectangular area around the center of the (full) ball and $C^0 = -1$ elsewhere, such that the evolving droplet has a given volume denoted by $V$. The initial velocity $\boldsymbol{u}^0$ is set to zero. A contact angle of $\theta_{eq} = 180°$ is fixed on the outer layer of voxels at the sphere, such that the spherical boundary is always repelling the droplet, while $\theta_{eq}$ on the bottom of the domain can be chosen arbitrarily. In the stationary state, a sessile droplet will evolve, which should reveal a contact angle that approximates the prescribed value of $\theta_{eq}$, cf. Fig. 5. In the sharp interface case, the radius $r_{sh}$ (sh for sharp) of the circular contact area between droplet and wall can be determined by the alge-

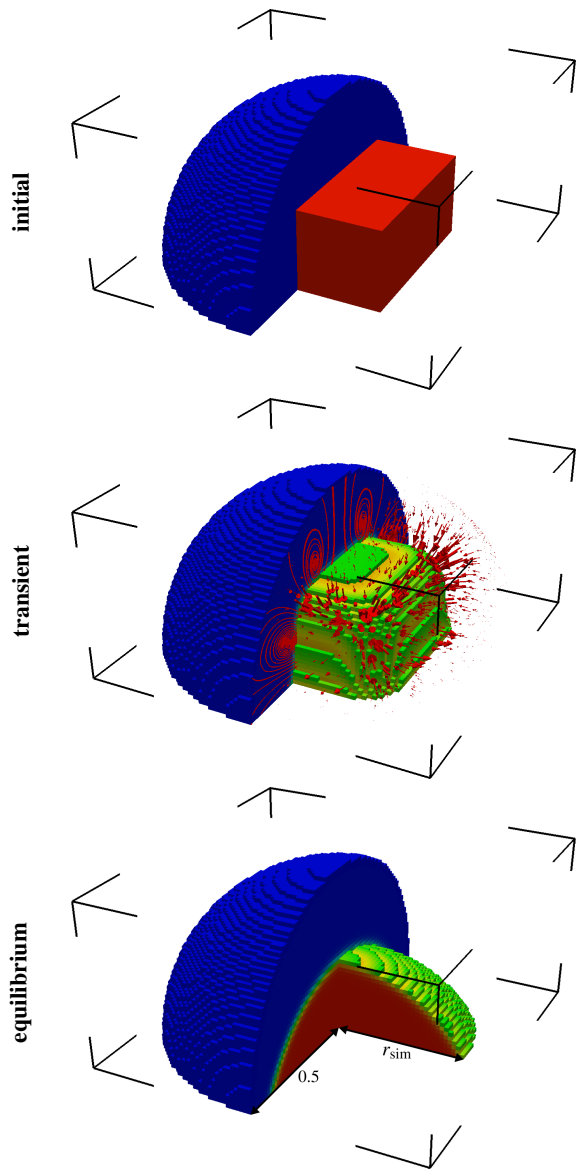

Figure 5: Initial configuration of phase distributions *(top)*, transient state with nonvanishing velocity *(center)*, and equilibrium configuration *(bottom)* for $\theta_{eq} = 60°$. Phase A with volume $V = 1/16$ is illustrated *red*, phase B in *blue*, and the interface center in *green*. The mesh width is $h = 1/80$ here.

## 5.1.2. Simulation data

We apply this numerical scenario to the non-rotated domain and to voxel set overlying half ball that are rotated by 45° in $x$ direction and a third one rotated by 45° in $x$ and in $y$ direction. We refer to these three cases as *planar*, *stair*, and *criss-cross* due to the voxel alignment at the bottom, cf. Fig. 4). In each of the considered cases, the vicinity of all faces $e_h$ that approximate the bottom of the half ball have the same local geometry and thus, each of those faces has the same correction factor $\delta_{e_h}$. We prescribe a droplet volume of $V = 1/16$ and contact angles of $\theta_{eq} = 60°, 90°, 120°$ using grid sizes and interface parameters $h, \epsilon = 1/20, 1/40, 1/80$ (obeying the constraint of $h \leq \epsilon$). For each set of parameters, the relative error between the radius $r_{sim}$ at stationary state (i. e. thermodynamic equilibrium) and the analytical radius $r_{sh}$ by (13) is computed. To measure $r_{sim}$, the values of $C$ are plotted against a line through the points $(0, 0, 0.5)$ and $(1, 1, 0.5)$ for the planar case, through $(0, 0.5, 0)$ and $(1, 0.5, 1)$ for the stair case, and through $(0.5, 0, 0)$ and $(0.5, 1, 1)$ for the criss-cross case.

## 5.1.3. Outcome

Results are listed in Tab. 1. For the planar case, the energy correction factor at bottom faces is $\delta_{e_h} = 1$, i.e., the original formulation is equivalent with the boundary condition for jagged surfaces (11). For angles $\theta_{eq}$ of 60°, 90°, and 120°, all relative errors are 1.3%, which indicates that the contact angle model (7) is accurate on planar surfaces. A prescribed contact angle of 90° (i. e., the CH equation in standard form since the right-hand sides of (7) and (11) vanish) is satisfied in the stair and the criss-cross case in a similar fashion as in the planar case.

However, for $\theta_{eq} = 60°$ on the stair and criss-cross geometry, we observe a much smaller contact angle (i. e. $r_{sim} \gg r_{sh}$) and for $\theta_{eq} = 120°$, we observe a larger much contact angle (i. e. $r_{sim} \ll r_{sh}$). For the stair case, where we have plain-ordered voxels in one direction and staged voxels in the other direction (cf. Fig. 4 (b)), it is remarkable that this amplification of the contact angle is of equal magnitude in *each* direction. This fact is a requirement for the contact angle correction approach, since the involved surface energy damping acts as scalar on each face of the grid and is thus isotropic. Using the corrected boundary condition (11) instead, the relative errors are reduced to an acceptable magnitude (cf. Fig. 3 (a–c) for the illustration of the three different geometries).

Note that solving the coupled CHNS system as well as solving the CH subsystem (3), (4) produces the same data listed in Tab. 1. Therefore, we restrict our computation to the latter case in Secs. 5.2 and 5.3.

| $\theta_{eq}$ | $r_{sh}$ | $\delta_{e_h}$ | planar | | stair | | criss-cross | |
|---|---|---|---|---|---|---|---|---|
| | | | $r_{sim}$ | $\frac{|r_{sh}-r_{sim}|}{r_{sh}}$ | $r_{sim}$ | $\frac{|r_{sh}-r_{sim}|}{r_{sh}}$ | $r_{sim}$ | $\frac{|r_{sh}-r_{sim}|}{r_{sh}}$ |
| **original** | | | | | | | | |
| 90° | 0.310 | | 0.306 | 1.3% | 0.306 | 1.3% | 0.307 | 1.0% |
| 60° | 0.396 | | 0.391 | 1.3% | 0.442 | **11.6%** | 0.473 | **19.4%** |
| 120° | 0.226 | | 0.223 | 1.3% | 0.180 | **20.4%** | 0.124 | **45.1%** |
| **corrected** | | | | | | | | |
| 90° | 0.310 | 1.000 | 0.306 | 1.3% | 0.306 | 1.3% | 0.307 | 1.0% |
| 60° | 0.396 | 0.707 | 0.391 | 1.3% | 0.389 | 1.8% | 0.390 | 1.5% |
| 120° | 0.226 | 0.577 | 0.223 | 1.3% | 0.229 | 1.3% | 0.226 | 0.0% |

Table 1: Measurements on planar, stair, and criss-cross surface without and with energy correction using $V = 1/16$, $h = \epsilon = 1/80$ ($r_{sim}$ is measured at $C = 0$, energy correction factors $\delta_{e_h}$ is global).

## 5.2. Droplet on a spherical surface

In this scenario, a droplet with radius $R$ and center $\boldsymbol{x}_c$ is placed next to the boundary of a domain $\partial\Omega_h$ with $h = 1/100$ that approximates a sphere (containing 47 160 faces at the solid–fluid interface). Compared to the setup of Sec. 5.1, the local geometry at the three-phase contact line is more complex, yielding different energy correction factors on the faces $e_h$. Let the initial order parameter be given by

$$C^0(\boldsymbol{x}) = \tanh\left(\frac{R - \|\boldsymbol{x} - \boldsymbol{x}_c\|}{\sqrt{2}\,\epsilon}\right) \qquad (14)$$

where $R := 0.5$, $\boldsymbol{x}_c := (0.5, 0.5, 0)$, and $\epsilon := h$. The interface profile of this initial droplet is in equilibrium state [37]. We prescribe a contact angle of $\theta_{eq} = 120°$ and drive the system to the stationary state.

Figure 6 shows the droplet in equilibrium state. Using the original boundary condition (7) yields an angle of 141°, while we measure an angle of 121° when the corrected boundary condition (11) is used.

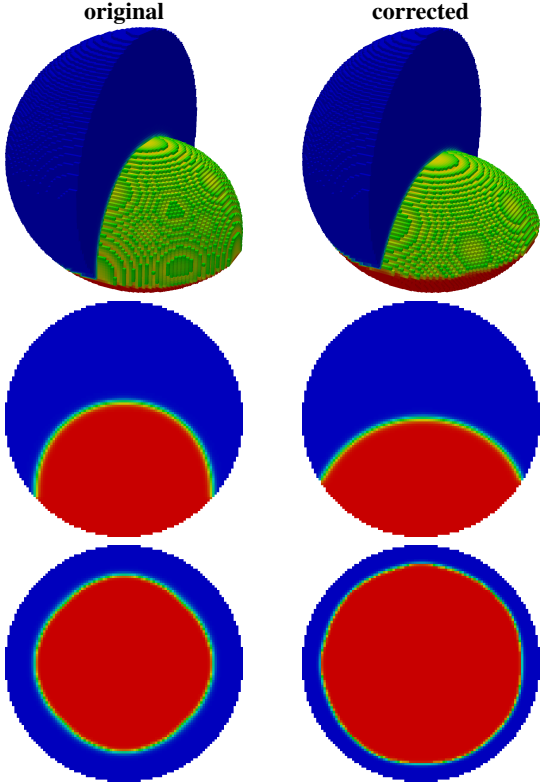

Figure 6: Droplet sitting on a spherical surface at equilibrium state with prescribed contact angle $\theta_{eq} = 120°$. The grid size is $h = 1/100$. 3D view with partially transparent phase B *(top)*, slice $x = 0.5$ through the domain *(center)*, and "bottom view" of the domain, i.e., view in $z$ direction without perspective *(bottom)*. The measured contact angle is 141° using (7) *(left)* and 121° using (11) *(right)*.

### 5.3. Droplet inside a pore cavity

While surface geometries are regular and artificial in Secs. 5.1 and 5.2 with regular geometries, in this scenario, the computational domain is taken from a µCT scan of a Castlegate Sandstone. A spherical droplet is placed in one of the pore cavities using (14) and the equilibrium configuration is computed using either of the contact angle conditions (7) or (11). At equilibrium, the contact angle distribution along the three-phase contact line is measured via the algorithm of Scanziani et al. [38]. This algorithm extracts 2D planes perpendicular to an estimated three-phase contact line and uses the physical constraint of constant curvature, given by Young–Laplace law at equilibrium, to fit a circle to the fluid–fluid interface. This way it is possible to estimate contact angles for arbitrary points along the estimated contact line.

Figure 7 shows the stationary state of the droplet with a prescribed contact angle $\theta_{eq}$ of 45°. Using the original contact angle boundary condition (7), wettability of the jagged surface is overestimated yielding contact angles along the three-phase contact line that are significantly smaller than the prescribed one. The corrected boundary condition (11) yields contact angles that are closer to the desired one. In fact, application of Scanziani's algorithm, the probability density function distribution illustrated in Fig. 8 is obtained, revealing a mean value of 35.7° at a standard deviation of 9.5°. The tuning parameters of the algorithm are kept as in [38].

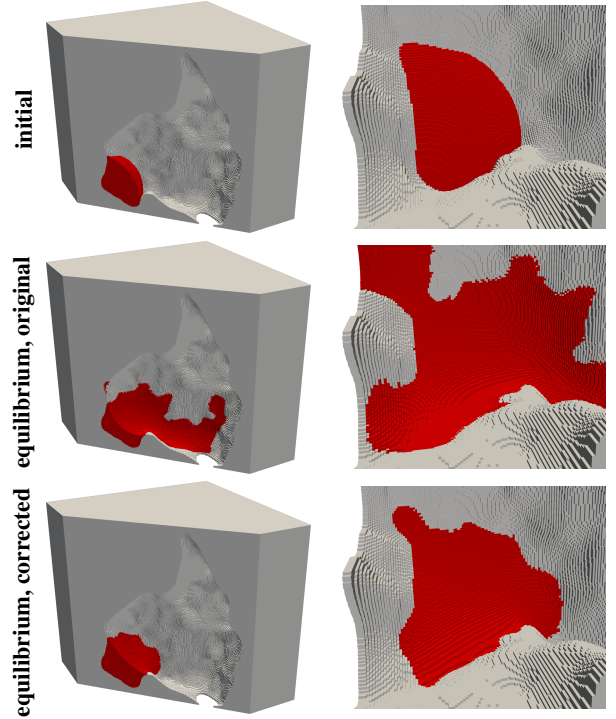

Figure 7: Droplet sitting in a pore cavity with prescribed contact angle $\theta_{eq} = 45°$ at initial state *(top)*, at equilibrium state using (7) *(center)*, and at equilibrium state using (11) *(bottom)*. The underlying µCT image data consists of 256 voxels in each direction. The rock is visualized *gray*, phase A *red*, and phase B *transparent*.

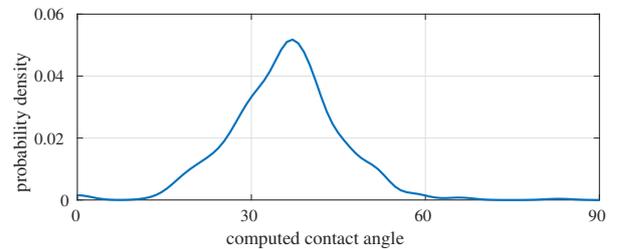

Figure 8: Probability density function of computed contact angles along the estimated three-phase contact line for the corrected scenario illustrated in Fig. 8.

## 6. Conclusion

Accurate modeling of wetting phenomena and its interactions with the multi-phase transport physics is of fundamental importance for the robust and reliable operation of a pore-scale flow simulator. We developed a novel energy-based correction technique for the correct implementation of the equilibrium wetting boundary condition within the phase-field framework for pore-scale image volumes with jagged surfaces. Such image-volumes typically stem from µCT imaging and the ensuing voxelization step. They are utilized as input for direct pore-scale flow simulators that circumvent the complex and computationally costly mesh-generation step.

The proposed technique is based on the correction of the surface energy as a function of the local geometry. In this context, we propose the introduction of surface energy correction factors for each fluid–solid voxel face that counterbalance the difference of the voxel-set surface area with the underlying smooth one. The local approximation of the fluid–solid surfaces renders the algorithm particularly efficient.

The proposed technique operates on the strong formulation of the model and can easily be implemented as a pre-processing step before running the pore-scale flow simulation. As such, this semi-analytical technique is independent of the discretization parameters and is generally applicable independent of the discretization method used to numerically solve the underlying phase-field equations.

We have systematically validated the equilibrium contact angle correction scheme on test cases with increasing complexity. We have also demonstrated its application on a real high-resolution image-volume of a Castlegate Sandstone sample. The proposed technique is simple, yet, robust and effective, therefore, it is implemented as an integral component of a phase-field method based pore-scale flow simulator.

## 7. Appendix

For completeness, we state the proof of Proposition 1 (see also [39, 40]): Differentiating (10) in time, using Young's equation (5), the boundary condition (7), and the definition of $\phi$ yields

$$\begin{aligned}
\mathrm{d}_t F &= \int_\Omega \rho\, \boldsymbol{u} \cdot \partial_t \boldsymbol{u} + \beta\, \Psi'(C)\, \partial_t C + \alpha\, \boldsymbol{\nabla} C \cdot \boldsymbol{\nabla}(\partial_t C) \\
&\quad + \int_{\partial\Omega} (\sigma_{\mathrm{Bs}} - \sigma_{\mathrm{As}})\, g'(C)\, \partial_t C \\
&= \int_\Omega \rho\, \boldsymbol{u} \cdot \partial_t \boldsymbol{u} + \beta\, \Psi'(C)\, \partial_t C - \alpha\, \Delta C\, \partial_t C \\
&\quad + \int_{\partial\Omega} \alpha\, (\boldsymbol{\nabla} C \cdot \boldsymbol{n})\, \partial_t C + \sigma_{\mathrm{AB}}\, \cos(\theta_{\mathrm{eq}})\, g'(C)\, \partial_t C \\
&= \int_\Omega \rho\, \boldsymbol{u} \cdot \partial_t \boldsymbol{u} + \phi\, \partial_t C\,.
\end{aligned}$$

Substituting $\partial_t C$ and $\partial_t \boldsymbol{u}$ from (4) and (9), respectively, yields

$$\begin{aligned}
\mathrm{d}_t F = \int_\Omega &-\rho\, \boldsymbol{u} \cdot (\boldsymbol{u} \cdot \boldsymbol{\nabla} \boldsymbol{u}) - \boldsymbol{u} \cdot \boldsymbol{\nabla} p + \mu\, \boldsymbol{u} \cdot \Delta \boldsymbol{u} \\
&+ \phi\, \boldsymbol{\nabla} \cdot (M\, \boldsymbol{\nabla} \phi) - C\, \boldsymbol{u} \cdot \boldsymbol{\nabla} \phi - \phi\, \boldsymbol{u} \cdot \boldsymbol{\nabla} C\,.
\end{aligned}$$

The following identities conclude the proof, where we used the incompressibility constraint (8), the no-slip boundary condition for $\boldsymbol{u}$, and the no-flux boundary condition for $\phi$:

First term:

$$\int_\Omega \boldsymbol{u} \cdot (\boldsymbol{u} \cdot \boldsymbol{\nabla} \boldsymbol{u}) = \int_\Omega \frac{1}{2} \boldsymbol{u} \cdot \boldsymbol{\nabla}(\boldsymbol{u} \cdot \boldsymbol{u}) - \boldsymbol{u} \cdot (\boldsymbol{u} \times \boldsymbol{\nabla} \times \boldsymbol{u})$$
$$= \frac{1}{2} \int_\Omega \boldsymbol{\nabla} \cdot (\boldsymbol{u} \cdot \boldsymbol{u}\, \boldsymbol{u}) = \frac{1}{2} \int_{\partial\Omega} \boldsymbol{u} \cdot \boldsymbol{u}\, \boldsymbol{u} \cdot \boldsymbol{n} = 0\,.$$

Second term:

$$\int_\Omega \boldsymbol{u} \cdot \boldsymbol{\nabla} p = \int_\Omega \boldsymbol{\nabla} \cdot (p\, \boldsymbol{u}) = \int_{\partial\Omega} p\, \boldsymbol{u} \cdot \boldsymbol{n} = 0\,.$$

Third term (sum over $i$):

$$\int_\Omega \boldsymbol{u} \cdot \Delta \boldsymbol{u} = \int_\Omega -\boldsymbol{\nabla} u^i \cdot \boldsymbol{\nabla} u^i + \boldsymbol{\nabla} \cdot (u^i\, \boldsymbol{\nabla} u^i)$$
$$= -\int_\Omega \boldsymbol{\nabla} u^i \cdot \boldsymbol{\nabla} u^i + \int_{\partial\Omega} u^i\, \boldsymbol{\nabla} u^i \cdot \boldsymbol{n} = -\int_\Omega \boldsymbol{\nabla} \boldsymbol{u} : \boldsymbol{\nabla} \boldsymbol{u}\,.$$

Fourth term:

$$\int_\Omega \phi\, \boldsymbol{\nabla} \cdot (M\, \boldsymbol{\nabla} \phi) = \int_\Omega -\boldsymbol{\nabla} \phi \cdot (M\, \boldsymbol{\nabla} \phi) + \boldsymbol{\nabla} \cdot (\phi\, M\, \boldsymbol{\nabla} \phi)$$
$$= -\int_\Omega \boldsymbol{\nabla} \phi \cdot (M\, \boldsymbol{\nabla} \phi) + \int_{\partial\Omega} \phi\, M\, \boldsymbol{\nabla} \phi \cdot \boldsymbol{n}\,.$$

Fifth and sixth term:

$$\int_\Omega \boldsymbol{u} \cdot \boldsymbol{\nabla}(C\, \phi) = \int_\Omega \boldsymbol{\nabla} \cdot (C\, \phi\, \boldsymbol{u}) = \int_{\partial\Omega} C\, \phi\, \boldsymbol{u} \cdot \boldsymbol{n} = 0\,.$$

## List of symbols

| Symbol | Unit | Description |
|---|---|---|
| $\alpha$ | J m$^{-1}$ | Factor in gradient energy density (constant), cf. (2). |
| $\beta$ | J m$^{-3}$ | Factor in chemical energy density (constant), cf. (2). |
| $C$ | – | Order parameter, difference of mass or volume fractions $C_A$, $C_B$, $C = C_A - C_B$, physically meaningful in $[-1, 1]$. |
| $\delta_{e_h}$ | – | Energy-correction factor, a positive constant on each voxel face $e_h \subset \partial\Omega_h$. |
| $\epsilon$ | m | Interface parameter (constant). |
| $F$ | J | Total mixture energy, functional in $C$ and $\boldsymbol{u}$, cf. (10). |
| $g$ | – | Blending function (function of $C$). |
| $h$ | m | Mesh size, edge length of one voxel. |
| $\epsilon$ | m | Interface parameter. |
| $\boldsymbol{j}_1, \boldsymbol{j}_2$ | m s$^{-1}$ | Flux due to diffusion and advection, cf. (4). |
| $J$ | m | Open time interval, $J := (0, t_{end})$. |
| $\mu$ | J s m$^{-3}$ | Dynamic viscosity of the mixture (constant). |
| $M$ | m$^5$ J$^{-1}$ s$^{-1}$ | Mobility of the mixture (constant). |
| $\boldsymbol{n}$ | – | Unit normal on $\partial\Omega$ outward of $\Omega$. |
| $\Omega$ | m$^3$ | Spatial domain (pore space) with smooth boundary of which $\Omega_h$ is a voxel approximation. |
| $\Omega_h$ | m$^3$ | Union of voxels overlaying the smooth domain $\Omega$. |
| $\partial\Omega$ | m$^2$ | Boundary of $\Omega$, fluid–solid interface. |
| $\partial\Omega_h$ | m$^2$ | Union of voxel faces approximating the fluid–solid interface $\partial\Omega$. |
| $p$ | J m$^{-3}$ | Reduced pressure, $p = P + \beta\Psi(C) + \frac{\alpha}{2}|\boldsymbol{\nabla} C|^2 + C\phi$ with pressure $P$, cf. [40]. |
| $\phi$ | J m$^{-3}$ | Pseudo chemical potential. |
| $\Psi$ | – | Chemical energy density factor (function of $C$). |
| $\rho$ | kg m$^{-3}$ | Mass density of the mixture (constant). |
| $\sigma_{AB}$ | J m$^{-2}$ | Surface tension between phase A and B (constant). |
| $\sigma_{As}$ | J m$^{-2}$ | Surface tension between phase A and solid phase (constant). |
| $t$ | s | Time variable, $t \in \overline{J}$. |
| $\theta_{eq}$ | | Equilibrium contact angle (constant). |
| $\boldsymbol{u}$ | m s$^{-1}$ | Advection velocity of the mixture, $\boldsymbol{u}: J \times \Omega \to \mathbb{R}^3$. |
| $\boldsymbol{x}$ | m | Space variable, $\boldsymbol{x} \in \overline{\Omega}$. |

Table 2: List of symbols.